\documentclass[twocolumn,showpacs,preprintnumbers,bibnotes]{revtex4}
\usepackage{graphicx}
\usepackage{dcolumn}
\usepackage{bm}
\usepackage{amssymb}
\usepackage{amsmath}
\usepackage{epsfig}
\usepackage{bm}

\begin{document}


\title{Observation of Rabi oscillation of light assisted by atomic spin wave}

\author{L. Q. Chen$^{1}$}
\author{Guo-Wan Zhang$^{1}$}
\author{Cheng-ling Bian$^{1}$}
\author{Chun-Hua Yuan$^{1}$}
\author{Z. Y. Ou$^{1,2,*}$}
\author{Weiping Zhang$^{1\dag}$}
\affiliation{$^{1}$State Key Laboratory of Precision Spectroscopy,
Department of Physics, East China Normal University, Shanghai
200062, P. R. China}
\affiliation{$^{2}$Department of Physics, Indiana University-Purdue University Indianapolis,
402 N. Blackford Street, Indianapolis, IN 46202, USA}

\date{\today }

\begin{abstract}
Coherent conversion between a Raman pump field and its Stokes field is observed in a Raman process with a strong atomic spin wave initially prepared by another Raman process operated in the stimulated emission regime. The oscillatory behavior resembles the Rabi oscillation in atomic population in a two-level atomic system driven by a strong light field. The Rabi-like oscillation frequency is found to be related to the strength of the pre-built atomic spin wave. High conversion efficiency of 40\% from the Raman pump field to the Stokes field is recorded and it is independent of the input Raman pump field. This process can act as a photon frequency multiplexer and may find wide applications in quantum information science.

\end{abstract}

\pacs{42.50.-p,42.50.Gy,42.50.Ex,42.65.Dr}
\maketitle

In quantum communication, photon is the basic information carrier. A single-photon usually carries limited amount of information in a single channel, although quantum entanglement may increase the information with dense-coding scheme \cite{mat,li}. Thus to increase the amount of  information, multi-channel transmission is required. In classical communication, frequency multiplexing method creates multiple independent channels for transmitting classical information. In quantum communication, we may use the similar technique but different channels may not be independent to each other\cite{zei}.

In many quantum information protocols, a single-photon state is the preferred quantum state. A frequency multiplexed entangled single-photon state has the form of (two frequency components)
\begin{eqnarray}
|1_{\omega_1,\omega_2}\rangle = c_1 |1\rangle_{\omega_1}|0\rangle_{\omega_2} + c_2|0\rangle_{\omega_1}|1\rangle_{\omega_2}.\label{1}
\end{eqnarray}
Creation of the above state requires coherent photonic frequency conversion, which converts photon from one frequency to another in a coherent and efficient way.

As is well-known, optical frequency conversion is routinely achieved in nonlinear optics \cite{shen} where high intensity is a prerequisite for efficient conversion. Recently, efficient photon frequency up-conversion is achieved by sum-frequency generation \cite{kum,de,kwi2,gsi} while frequency down-conversion is suggested in simulated parametric process \cite{ou08}. However, all these are based on $\chi^{(2)}$-materials and require a strong driving field simultaneously present to provide the nonlinear interaction. The huge background from the strong driving field, on the other hand, is troublesome for the detection of the weak quantum field.

Besides the $\chi^{(2)}$ process for frequency conversion, the effect of electromagetically induced transmission (EIT) was used by Jain et al \cite{ja,ja2} to convert efficiently light pulses of 450 nm to 290 nm in a Raman process involving atomic coherence.  More recently, it was suggested that such a Raman process can be used to convert quantum state from light to atom and vice versa \cite{ou08}. This is based on a strong injection of Stokes field and stimulated emission. Efficient frequency conversion of weak fields in EIT assisted four-wave mixing was also reported \cite{zib}.

In this letter, we propose and experimentally demonstrate a new scheme to realize the coherent frequency conversion of photons. This scheme is based on the coherent coupling of photons to pre-built atomic spin waves in an atomic ensemble through a Raman scattering process. Here, the pre-built atomic spin waves play the role of strong coupling ``field" which drives the photons to coherently oscillate in the pump mode and Stokes mode in the Raman process. This effect is quite similar to the Rabi oscillation between two atomic states in a two-level system driven by a strong light field \cite{rabi}. In our scheme, the two modes of pump and Stokes photon have the equivalent roles to the two atomic states and the pre-built atomic spin waves replace the role of the strong driving light field coupling the two states in the traditional atomic Rabi oscillation.

\begin{figure}
\begin{center}
\includegraphics[angle=-90,bb=100 60 330
1000,scale=0.33]{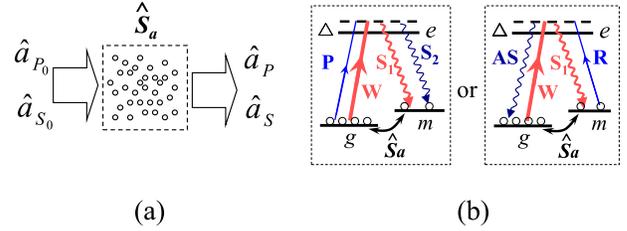}
\end{center}
\caption{(Color online) (a) Raman process with a strong pre-built atomic spin wave $\hat S_a$ (dashed box) for efficient conversion between the pump and the Stokes fields. (b) Schematic atomic levels showing the coupling between light and atom.} \label{f1}
\end{figure}
Consider a Raman process, shown in Fig.1,  with an ensemble of $N_a$ atoms with a pair of lower
level meta stable states $|g\rangle, |m\rangle$, the collective
atomic spin field $\hat S_a\equiv
(1/\sqrt{N_a})\sum_i|g\rangle_i\langle m|$ is coupled to the
Stokes and the Raman pump field via an upper excited level $e$ by the Hamiltonian \cite{DLCZ}:
\begin{eqnarray}
\hat H_R=i\hbar \eta \hat a_{P} \hat a_{S}^{\dag}\hat S_a^{\dag}
-i\hbar \eta \hat a_S\hat S_a\hat a_{P}^{\dag},\label{Hr}
\end{eqnarray}
where $\eta = g_{eg}g_{em}/\Delta$ with $g_{eg},g_{em}$ as the coupling coefficients between the excited state and the lower level states. $\Delta$ is the detuning from the excited state for  both the Stokes and Raman pump fields, which satisfy the two-photon resonance condition: $\omega_P-\omega_S = \omega_{mg}$.

For the Hamiltonian in Eq.(\ref{Hr}), we consider a situation when the atomic spin wave is strong and can be treated as a classical wave: $\hat S_a \rightarrow {\cal S}_a$. Then the Hamiltonian in Eq.(\ref{Hr}) is changed to
\begin{eqnarray}
\hat H_{PS}=i\hbar \eta {\cal S}_a^* \hat a_{P} \hat a_{S}^{\dag}
-i\hbar \eta {\cal S}_a \hat a_S\hat a_{P}^{\dag}.\label{Hws}
\end{eqnarray}
In this case, the Raman pump field $\hat a_{P}$ is the relatively weak quantum field and renamed as the Raman probe field. We use the same symbol $P$ for it. The operator evolution for $\hat H_{PS}$ is \cite{lou}
\begin{eqnarray}
&&\hat a_{S} = \hat a_{S0}\cos\Omega t + \hat a_{P0}\sin\Omega t\cr
&&\hat a_{P} = \hat a_{P0}\cos\Omega t - \hat a_{S0}\sin\Omega t,\label{aw}
\end{eqnarray}
where $\Omega = |\eta {\cal S}_a|$ and $\hat a_{P0},\hat a_{S0}$ are input Raman probe and Stokes fields, respectively. Eq.(\ref{aw}) is in the same form as the evolution for a beam splitter with $\hat a_{P0},\hat a_{S0}$ as the input and $\hat a_{P},\hat a_{S}$ as the output \cite{camp}. Notice that Eq.(\ref{aw}) is for quantum fields with any quantum states including single-photon state. In the case of a single-photon state input $|1_{P0}\rangle$ and  $\Omega t  = \pi/4$, the output state will be a frequency entangled single-photon state in Eq.(\ref{1}).

From Eq.(\ref{aw}), we find that there is an oscillation between the Raman probe field and the Stokes field as time progresses. The oscillation frequency is proportional to the amplitude of the atomic spin wave ${\cal S}_a$. This is in exactly the same form as the so-called Rabi frequency for the Rabi oscillation of atomic population in a system involving interaction between light and a two-level atom \cite{rabi}. It should be pointed out that, although the phenomenon is also called Rabi oscillation and involves two light fields and a similar atomic level structure as the two-photon Rabi oscillation \cite{loy,hick}, it is a completely different phenomenon. Two-photon Rabi oscillation, like conventional Rabi oscillation, is a coherent population oscillation between two atomic states driven by two light fields whereas here the oscillation is between two photonic states of different frequencies and the driving field is the strong atomic spin wave pre-built before the input of the light field.

Eq.(\ref{aw}) is a result without considering the propagation of light fields. This is valid in our case since the length of light pulses is much longer than the atomic medium.

Traditionally, the atomic spin wave can be prepared by EIT effect, that is, using two fields respectively on resonance with transitions $e\rightarrow g,m$ to convert optical coherence into atomic coherence between $g, m$ \cite{ja,ja2}. Recently, we demonstrated another method \cite{chen09} in which a large atomic spin wave amplitude can be produced in a Raman scattering process in the high gain regime with a strong Raman pump field.
The process can also be described by the Hamiltonian in Eq.(\ref{Hr}) but with a strong Raman pump field. We will label this strong Raman pump field as the ``write" field because it writes a strong atomic spin wave. In this case, we can replace operator $\hat a_{P}$ in Eq.(\ref{Hr}) with a classical amplitude $A_W$ and we have
\begin{eqnarray}
\hat H_W=i\hbar \eta A_{W} \hat a_{S}^{\dag}\hat S_a^{\dag}
-i\hbar \eta \hat a_S\hat S_aA_{W}^{*},\label{HW}
\end{eqnarray}
which leads to \cite{lou}
\begin{eqnarray}
\hat S_{a} = \hat S_{a0}\cosh\kappa t +\hat a_{S0}^{\dag}\sinh\kappa t,\label{Sa}
\end{eqnarray}
where $\kappa = |\eta A_W|$. For atoms initially in the ground state $g$ and the Stokes field in vacuum, we have $\langle \hat S_a^{\dag}\hat S_a\rangle = (\sinh \kappa t)^2$. A significantly large spin wave amplitude ${\cal S}_a$ is produced with $|{\cal S}_a|^2 \equiv \langle \hat S_a^{\dag}\hat S_a\rangle\sim \sinh ^2(|\eta A_W t|)$ for Eq.(\ref{aw}). Although only one laser field is present here, i.e., the ``write" field, as compared to the two fields needed in the EIT scheme \cite{ja,ja2}, the scheme involves, in essence, two fields. The second field is the spontaneously generated Stokes field ($\hat a_S$). In our experimental demonstration described next, we adopt this simpler method for the creation of a strong atomic spin wave.
\begin{figure}
\begin{center}
\includegraphics[angle=-90,bb=80 100 380
1000,scale=0.38]{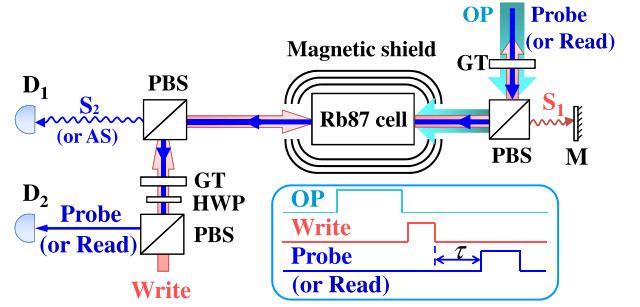}
\end{center}
\caption{(Color online) Experimental layout. Inset: time sequence for different fields. OP: optical pumping pulse; Write: spin wave creation pulse; Probe or Read: fields to be converted. GT, Glen-Thompson prism; HWP: half wave plate. } \label{f2}
\end{figure}

The sketch of the experimental arrangement is shown in Fig.2. The atomic medium is a 75mm long isotopically enriched rubidium-87 cell inside three layers of magnetic shield. The cell is heated to 82-86$^{\circ}$C with a bi-filar resistive heater. The two lower level energy states $(g,m)$ are the hyperfine ground states of $5^2S_{1/2}(F = 1; 2)$. The upper excited level (level $e$) is $5^2P_{1/2}$. In order to have the spin wave picture described above working, the atoms are required to be initially in the ground state $g$. This is achieved by optical pumping with a laser (780 nm, line width $<$ 0.15MHz) tuned to $5^2S_{1/2}(F = 2) \rightarrow 5^2P_{3/2}$ transition (labeled as $OP$ in Fig.2). The depopulation degree of the $m$ state ($5^2S_{1/2}, F=2$) is measured to be smaller than 2.4\%. In our experiment, the initial atomic spin wave between states $g, m$ is created by the Raman process from a strong write field $W$ in the high gain regime. The write laser (795 nm, line width $<$ 0.3 MHz) is blue-detuned from the transition  $5^2S_{1/2}(F = 1) \rightarrow 5^2P_{1/2}$ by $\Delta = 1 - 2$ GHz, depending on the temperature of the cell and the power of the write laser, for maximum Raman scattering. This process is described by Eqs.(\ref{HW},\ref{Sa}). As seen in the time sequence in the inset of Fig.2, after the state initialization pulse $OP$, we send in a strong Raman write field $W$ (48 mW) to build up a strong initial atomic spin wave ${\cal S}_a$. Then after a short delay of $\tau (< 100 ns)$, we send a weak Raman probe pulse  $(P, power = 10 \mu W - 3 mW)$ in the opposite direction of $W$ for efficient conversion to the Stokes field. This probe pulse is very similar to the write pulse in that it is also blue-detuned from the transition  $5^2S_{1/2}(F = 1) \rightarrow 5^2P_{1/2}$ by $\Delta = 1 - 2$ GHz but it is from another laser (795 nm, line width $<$ 0.15 MHz). Both the write pulse ($W$) and the probe pulse ($P$) will produce Stokes fields labeled as $S_1, S_2$, respectively. For the reverse process of converting the Stokes field to the probe field, we inject light at the Stokes field in stead of the probe field. This is done by tuning the frequency of the laser that was originally the probe field close to the transition  $5^2S_{1/2}(F = 1) \rightarrow 5^2P_{1/2}$ with $\Delta ' = 1 - 2$ GHz. To make it different from the original probe, we call it the read pulse ($R$ in Figs.1,2). As a matter of fact, this process corresponds to the anti-Stokes process in a traditional CARS process \cite{dru}, in which the read field ($R$ in Figs.1,2) is tuned close to the Stokes transition ($S$ in Fig.1) and the converted probe ($P$) is the anti-Stokes field ($AS$ in Fig.2). The efficient conversion process is described by Eqs.(\ref{Hws},\ref{aw}).   The probe or the read pulse propagates in the opposite direction of the $W$ pulse. The counter-propagation arrangement achieves a better conversion than the co-propagation arrangement \cite{chen09}. In addition, we find that if we inject the generated Stokes $S_1$ from the spin wave creation pulse $W$ back into the atomic cell with a mirror $M$, it can greatly enhance the generation of the anti-Stokes ($AS$) by the read pulse $R$. Since the Stokes or anti-Stokes field has an orthogonal polarization from the Raman pump field (now called $W,P$) or the read, we can use a polarization beam splitter (PBS) to easily separate the Stokes  (anti-Stokes) and the write (read) fields and send them to separate detectors $(D_1,D_2)$, respectively for monitoring.

\begin{figure}
\begin{center}
\includegraphics[width=7cm]{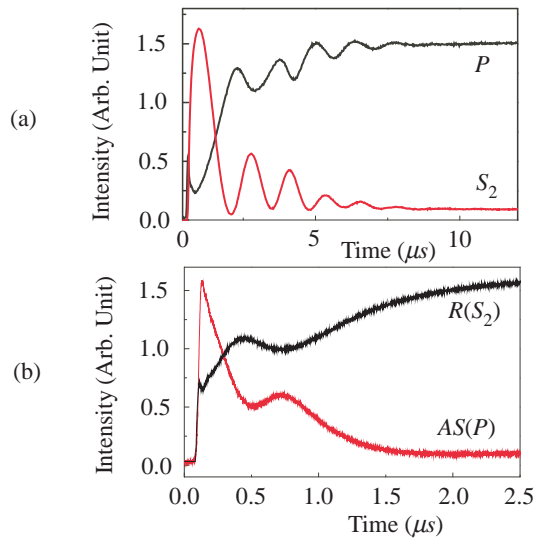}
\end{center}
\caption{(Color online) Observed Rabi oscillation (a) between the Stokes ($S_2$, red) and probe ($P$, black) fields and (b) between the anti-Stokes ($P$, red) and the read ($S_2$, black) fields.} \label{f3}
\end{figure}

Fig.3 gives the main result of our experiment. The generated Stokes (anti-Stokes) field shows an oscillatory behavior, which is complemented by the pass-through probe (read) field: they are 180$^{\circ}$ out of phase, indicating that the energy is transferred back and forth between them coherently. Apart from the decay of the signals, the observed oscillation is consistent with the sinusoidal behavior predicted in Eq.(\ref{aw}) for a Rabi oscillation between different frequencies of photons. The decay of the oscillation is due to decoherence in atomic spin wave, which typically gives a decay time of a few $\mu s$, due to atomic transit time in the interaction region. The anti-Stokes process shown in Fig.3(b) decays faster than the Stokes process in Fig.3(a) so that fewer periods of oscillation are observed. The faster decay is because the anti-Stokes process is initiated from the $m$ state which has a much smaller population than the $g$ state \cite{es}.

Next, we measure the frequency of the Rabi oscillation while the intensity of the spin wave creation write field ($W$) is varied. According to the theory presented earlier in Eqs.(\ref{aw},\ref{Sa}), the Rabi frequency depends on the strength of the initially prepared atomic spin wave ${\cal S}_a$, which in turn depends on the strength of the write field:
\begin{eqnarray}
\Omega \propto \sinh (|\eta A_{W}| t)\sim e^{|\eta A_{W}| t} ~~{\rm for~}|\eta A_{W}| t >>1,\label{Ome}
\end{eqnarray}
which leads to $\log (\Omega) \propto |A_{W}| \propto \sqrt{P_{W}}$ at high gain regime of $|\eta A_{W}| t>>1$. Fig.4 plots $\log (\Omega)$ as a function of square root of the power $P_{W}$ of the spin wave creation laser. The solid line is a linear fit to the experimental data (diamonds). It can be seen that the data follows very well a linear dependence. The flattening at high power end is perhaps because the maximum value of the atomic spin wave is reached due to limited number of available atoms.
There is a similar dependence of the Rabi frequency on the power of the write field ($W$) for the anti-Stokes process.

\begin{figure}
\begin{center}
\includegraphics[width=7cm]{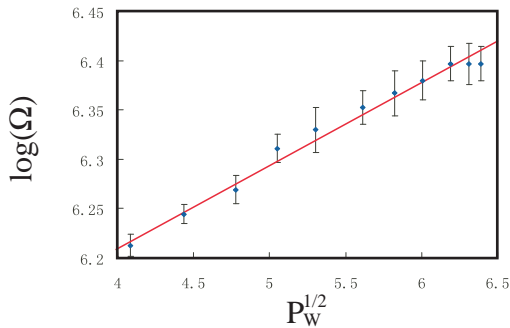}
\end{center}
\caption{(Color online) Logarithm of the Rabi frequency $\Omega$ versus the square root of the power of the write laser ($W$). } \label{f4}
\end{figure}

\begin{figure}
\begin{center}
\includegraphics[width=7cm]{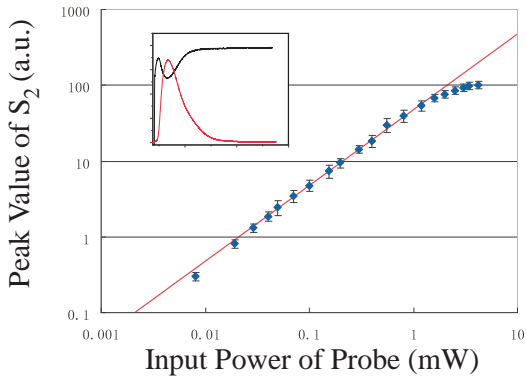}
\end{center}
\caption{(Color online) The peak value of $S_2$ versus the input power of $P$ in logarithmic scale. The red line has a slope of one. Inset: similar to Fig.3 but without oscillation.} \label{f5}
\end{figure}
From Eq.(\ref{aw}), we find that the frequency conversion process is a linear transformation process. To prove this,  we vary the input power of the Raman probe field ($P$) in a range that runs nearly three orders of magnitudes. To make it clear, we control the power of the spin wave creation field ($W_, P_{W} = 0.5 mW$) in such a way that there is only one peak in the Rabi oscillation of the generated Stokes field ($S_2$), as shown in the inset of Fig.5. This will produce the maximum conversion efficiency of about 40\% in pulse area.   In Fig.5, we plot the  peak value of the Stokes field ($S_2$) as a function of the input power of the Raman probe field ($P$) in logarithmic scales. Most of the data follows very well the solid red line, which has a slope of one, indicating a linear dependence. The points that are off the line at the higher end is because saturation is reached due to finite size of the prepared atomic spin wave while at the lower end, the sensitivity limit of the power meter is reached so that the measured power of the write field is not accurate. In principle, there is no lower limit for this linearity and thus we expect it will apply at single-photon level. The less-than-unit efficiency is probably due to the spatial mode mismatch between the initially prepared atomic spin wave (from the write field $W$) and the input probe field ($P$).

In summary, we have demonstrated quantum manipulation of photonic state via a ``field" of pre-built atomic spin wave in atomic medium. In this work, the roles of atom and photon are switched, which conceptually opens a new window for quantum simulation of atoms with photons and vice versa. Although in our experiment the creation of the atomic spin wave requires a strong write field,  the conversion process, in principle, does not directly involve a strong light field, in contrast to the high intensity required for most of the nonlinear optical processes. In fact, the initial atomic spin wave
can be prepared by other means such as a microwave pulse tuned to the
transition between states $|g\rangle$ and $|m\rangle$. Benefiting from the advantage of low detection noises without the background of the strong light, the frequency converter demonstrated here via atomic spin wave works at single-photon level. So it can find potential applications in single-photon wavelength division multiplexing to increase the channel capacity of quantum communication.

At first glance, the Rabi oscillation observed here is similar to the frequency beat effect routinely observed in CARS experiments \cite{lang,pes} in that sinusoidal oscillation in time is observed. However, there is a fundamental difference. The latter is an interference effect where energy is redistributed in time whereas the former is energy conversion between light of different frequencies.

This work is supported by the National Natural Science Foundation of China
under Grant Nos. 10828408 and 10588402, the National Basic Research
Program of China (973 Program) under Grant No. 2006CB921104,  the Program of Shanghai Subject Chief Scientist under Grant No. 08XD14017, Shanghai outstanding young teacher fund.
\newline
Email:$^*$zou@iupui.edu; $^\dag$wpzhang@phy.ecnu.edu.cn

\end{document}